\documentclass{article}

\PassOptionsToPackage{numbers, compress}{natbib}

\usepackage[preprint]{neurips_2021}

\usepackage[utf8]{inputenc} 	
\usepackage[T1]{fontenc}    	
\usepackage{hyperref}       	
\usepackage{url}            		
\usepackage{booktabs}       	
\usepackage{amsfonts}       	
\usepackage{nicefrac}       	
\usepackage{microtype}      	
\usepackage{xcolor}         		
\usepackage{graphicx}
\usepackage[export]{adjustbox}
\usepackage{comment}

\title{Rich dynamics caused by known biological brain network features resulting in stateful networks.}

\author{%
  Udaya B. Rongala\\
  Department of Experimental Medical Science\\
  Faculty of Medicine, Lund University\\
  Lund. Sweden.\\
  \texttt{udaya\_bhaskar.rongala@med.lu.se} \\
  \And
  Henrik J\"{o}rntell \\
  Department of Experimental Medical Science\\
  Faculty of Medicine, Lund University\\
  Lund. Sweden.\\
  \texttt{henrik.jorntell@med.lu.se} \\
}

\begin{document}

\maketitle

\begin{abstract}
  The mammalian brain could contain dense and sparse network connectivity structures, including both excitatory and inhibitory neurons, but is without any clearly defined output layer. The neurons have time constants, which mean that the integrated network structure has state memory. The network structure contains complex mutual interactions between the neurons under different conditions, which depend on the internal state of the network. The internal state can be defined as the distribution of activity across all individual neurons across the network. Therefore, the state of a neuron/network becomes a defining factor for how information is represented within the network. Towards this study, we constructed a fully connected (with dense/sparse coding strategies) recurrent network comprising of both excitatory and inhibitory neurons, driven by pseudo-random inputs of varying frequencies. In this study we assessed the impact of varying specific intrinsic parameters of the neurons that enriched network state dynamics, such as initial neuron activity, amount of inhibition in combination with thresholded neurons and conduction delays. The impact was assessed by quantifying the changes in mutual interactions between the neurons within the network for each given input. We found such effects were more profound in sparsely connected networks than in densely connected networks. However, also densely connected networks could make use of such dynamic changes in the mutual interactions between neurons, as a given input could induce multiple different network states.
\end{abstract}

\section{Introduction}
Recurrent excitatory and inhibitory loops are known to exist in mammalian central nervous system \cite{binzegger2004, song2005, koestinger2018, kar2021,zhu2000,douglas2009,obermayer2018,jorntell2003,pi2013,sultan2018}. These loops are known to be more complex than in traditional artificial neural networks. Furthermore, brain networks (neuronal networks) need not necessarily contain input and output neurons/layers, allowing every neuron in the network to contribute as an input and output function. The neurons were observed to surmise together to generate predictive models that helps in forming internal representing based on the external world inputs\cite{kawato1993,kawato1987,bastos2012,saleem2013,clark2013}. This theory makes every neuron within the brain networks an important component in representing the information, and hence the “networks internal state\footnote{Internal state of a network can be defined as distribution of activity across all individual neurons across the network.}” will be a defining factor in information representation. 
With this line of thinking, in this article we have explore the effect of input/neuron/network parameter on the internal state of the network. Towards this we have constructed a recurrent network comprising of both excitatory (ENs) and inhibitory neurons (INs) coupled using dense and sparse connectivity. We have chosen a fully connected network to have minimal assumptions. Both the ENs and INs are modelled based on similar neuron properties. A study conducted by Anton et. al. \cite{spanne2015}, based on a range of neuron types in the brain (\emph{in vivo}), they indicated that a neuron spike activity is a probability density function of the neuron membrane potential. Therefore, we have used a simple non-spiking neuron model that is computationally efficient and incorporating fundamental properties of a conductance-based Hodgkin–Huxley model \cite{rongala2021}. Such a neuron model will allow us to have minimal information loss withing the neurons in a network. We assessed the impact of varying specific intrinsic parameters of the neurons that enriched network state dynamics, such as input noise, initial neuron activity, amount of inhibition in combination with thresholded neurons and conduction delays.
In biology, the \emph{state of a network} was generally depicted as up and down states, which were characterized by high and low firing rates in the central nervous system \cite{steriade1993,cowan1994,lampl1999} and the advantages of such dynamic states was well studied \cite{parga2007,holcman2006,destexhe2009,sanchez2000}. The effect of network configurations such as conduction delays, and their effect on network states was also previously studied \cite{esir2018,woodman2011,ermentrout1998}. Whereas in this article we try to present the state of a network as a dynamically evolving entity, that is a function of the intrinsic parameters and resulting co-activation of all the neurons present in the network. We try to create this understanding, by addressing how intrinsic parameters of a neuron/network could affect the neurons activity, and further having its influence on the other neurons within the network.

\section{Methods}
\label{sec:methods}
\subsection{Neuron Model}

In this study we use a non-spiking Linear Summation neuron Model (LSM). LSM was a computationally compact neuron model that was designed to capture the important characteristic in a H-H conductance-based model \cite{rongala2021}. In LSM, the neuron membrane potential ($A$, Equation 1) was given by summation of weighted (\emph{w}) input synaptic activity (\emph{a}), that was normalized using static leak ($k_{static}$) and dynamic leak ($\tau_{dyn}$) components. The static leak was given by the total number of synapses on the neuron, reflecting the total number of open ion channels contributing to the membrane potential. The dynamic leak component mimics the effect of an RC circuit, reflecting the ion channels and membrane capacitance of a neuron. In the LSM adopted within this study, the neuron membrane activity was thresholded at zero resting potential (Equation 2). The continuous output neuron activity of LSM was intended to reflect the mean firing rate of a traditional spiking neuron model. The total neuron activity of LSM is given by following equations \ref{eqn1} \& \ref{eqn2},

\begin{equation} \label{eqn1}
\tau_{dyn}*\frac{dA}{dt} = -A(t) + \frac{\sum(w_{i}*a_{i}(t))}{k_{static} + \sum\left | w_{i}*a_{i}(t)\right |}
\end{equation}

\begin{equation} \label{eqn2}
if A < 0, then A = 0
\end{equation}

\subsection{Network Connectivity}
\begin{figure} 
  \centering
  \includegraphics[width=\textwidth]{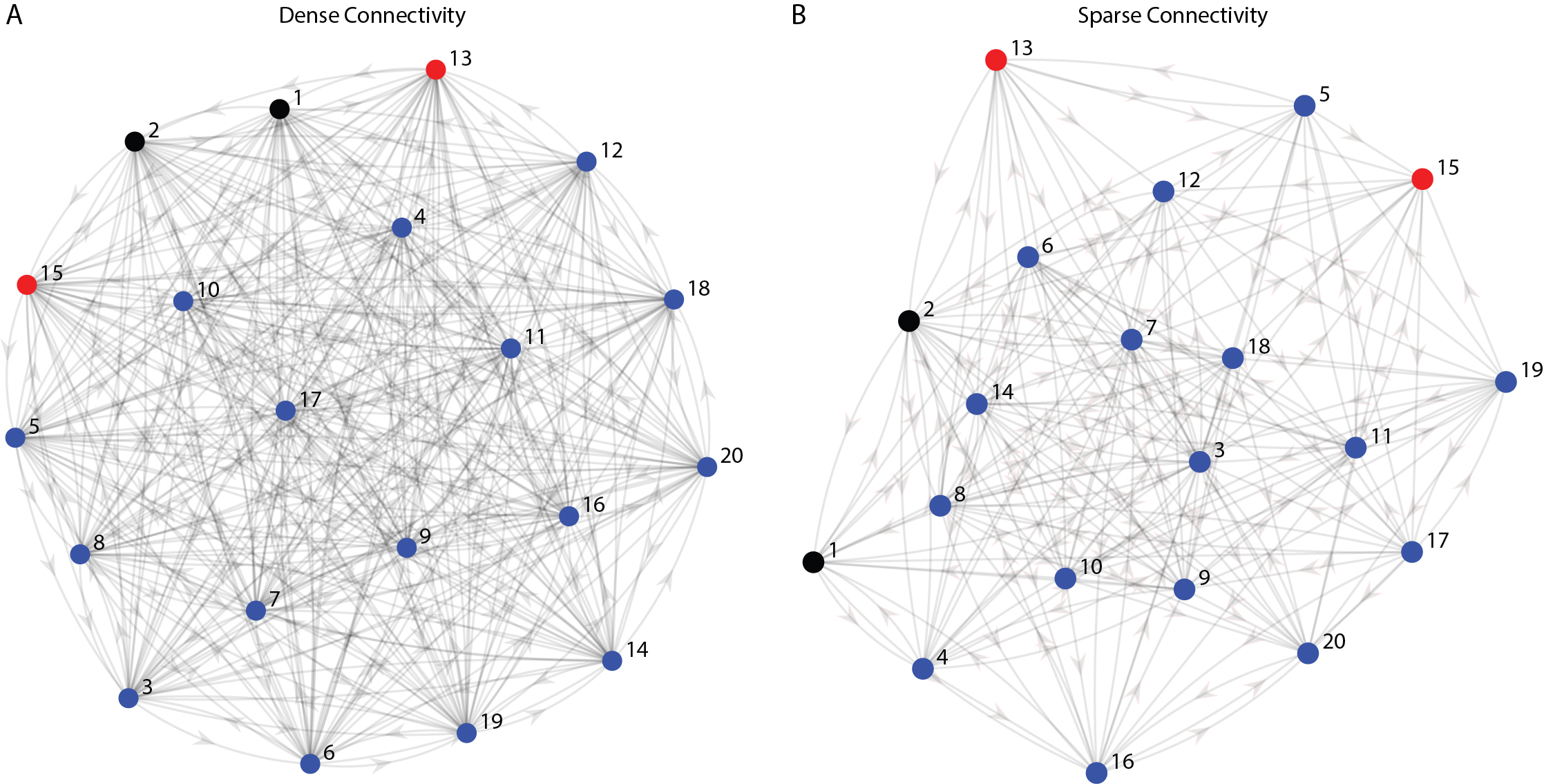}
  \caption{The networks that were studied. The networks comprised both excitatory (ENs, blue markers) and inhibitory neurons (INs, red marker). Both the networks consisted of a total of 20 neurons and two of the ENs (Neuron \#1, \#2) received the external inputs (black markers). (\textbf{A}) The network with dense connectivity, where all the neurons in the network were fully connected. (\textbf{B}) The network with sparse connectivity where 50\% of the connections were pseudo-randomly removed from the densely connected network.}
  \label{fig1}
\end{figure}

In this study we consider dense and sparse network connectivity configurations, comprising of both excitatory (\emph{ENs}, blue marker) and inhibitory neurons (\emph{INs}, red markers). In dense connectivity, we have a fully connected network where all the nodes within the network were bi-directionally connected (Figure \ref{fig1}A). In sparse connectivity, pseudo-randomly chosen 50\% of connections within the fully connected network were removed (Figure \ref{fig1}B). In this study, we considered a total of 20 neurons (including ENs and INs), where two of the ENs (Neuron \#1, \#2) received the inputs (Figure \ref{fig1}A, B, black markers). We explored different percentage of inhibitory neurons in the network, during which we switched a neuron from excitatory to inhibitory without any change in the connectivity. When a neuron was shifted from being excitatory to inhibitory, the weights of all outgoing connections for that neuron were converted from positive to negative weights while keeping the same value for the weight. While scaling the percentage of inhibitory neurons in the network, the index of INs were chosen pseudo-randomly.

\subsection{Conduction Delays}
The network dynamics in the presence of conduction delays (axonal delays between any two neurons in the network) was also explored in this study (Figure \ref{fig5}). The conduction delays were generated as normal distribution with a distribution mean ($\mu$) of 20 and standard deviation ($\sigma$) of 10. These conduction delays were fixed and not varied for different network simulation in this study. In simulations the value of conduction delay denotes the number of timesteps, that a signal is lagged between two neurons.

\subsection{Pseudo-Random Inputs}
The external inputs to the network, we generated pseudo-random spike trains of two different frequencies (5 and 50 Hz) with uniform normal distribution, using an inbuilt MATLAB function \emph{“randi”}. Further, these spikes were convoluted to resemble post-synaptic potentials using the following kernel equation \cite{mazzoni2008},

\begin{equation} \label{eqn3}
a_{i} = \frac{\tau_{m}}{\tau_{d}-\tau_{r}}*\\
\left [ exp\left ( \frac{-t-\tau_{l}-t^{*}}{\tau_{d}} \right ) 
 - exp\left ( \frac{-t-\tau_{l}-t^{*}}{\tau_{r}} \right )\right ]
\end{equation}

Where, $t^{*}$ is the input spike time, $\tau_{d}$ is the decay time (4 ms), $\tau_{r}$ is the rise time (12.5 ms), $\tau_{m}$ is the constant to calculate ratio between rise and decay times (21.3 ms) and $\tau_{l}$ is the latency time which is zero in this case. These values were chosen based on the previous work from \cite{rongala2018}.

\subsection{Noise Input}
A random gaussian noise was added to the convoluted signal  $(a_{i})$, as input to the neuronal networks. The noise was generated as uniform distribution ranging between 0 and a given maximum noise value (noise, Figures \ref{fig3} \& \ref{fig4}). Further to analyze the effect of noise on the network dynamics, we have also explored different levels of input noise, ranging between 0 and 0.2 with a step of 0.05 (Figures \ref{fig3} \& \ref{fig4}). We have generated 25 uniform distribution for each noise level (25 repetitions) to assess the variance in responses for a given network/input configuration.

\subsection{Initial neuron activity}
The network dynamics were explored based on the initial activity of the individual LSM neuron within the network. The initial activity was generated as uniform distribution ranging between 0 and a given maximum value (initial neuron activity, Figures \ref{fig3} \& \ref{fig4}), across all the neurons in the network. We have explored different level of the initial neuron activity, ranging between 0 and 0.2 with a step of 0.05 (Figures \ref{fig3} \& \ref{fig4}). We have generated 25 uniform distribution for each given initial neuron activity (25 repetitions) to assess the variance in responses for a given network/input configuration.

\subsection{Synaptic Weights}
All connection in the network were weighted. The weights were generated as normal distribution with a distribution mean $(\mu)$ of 0.3 and standard deviation $(\sigma)$ of 0.1. The weights of INs outgoing connections were set negative and the ENs were made positive.

\subsection{Statistical Analysis}
\subsubsection{Cross correlation}
The correlation index measure was used to compute the similarity between all the neuron output responses (Figure \ref{fig3}). The correlation index was computed across all the 20 neuron pairs, and the average across all pairs was reported. This measure was computed using an inbuilt MATLAB function \emph{“xcorr”} (with zero lag), which produces values (cross-correlation index) from 0 (uncorrelated) to 1 (similar).

\subsubsection{Covariance}
The covariance measure was used to determine how the neurons co-activate for a given input. The covariance measure was computed across all the 20 neurons activity. We have used an inbuilt MATLAB function \emph{“pca”} and extracted the covariances between all the 20 neurons for first principal component (explained > 90\% of information in the network). The \emph{pca} was computed on the entire time window (1s) of neuron output responses for a given input. To assess the effect of a given network configuration (input noise, INA, Inhibition), we have computed the Euclidean distance as a function of covariance matrix, this was reported as “covariance index”. The value of covariance index at zero indicate high similarity between the neuron covariances for given network configurations.

\begin{figure}[ht] 
  \centering
  \includegraphics[width=0.5\textwidth, center]{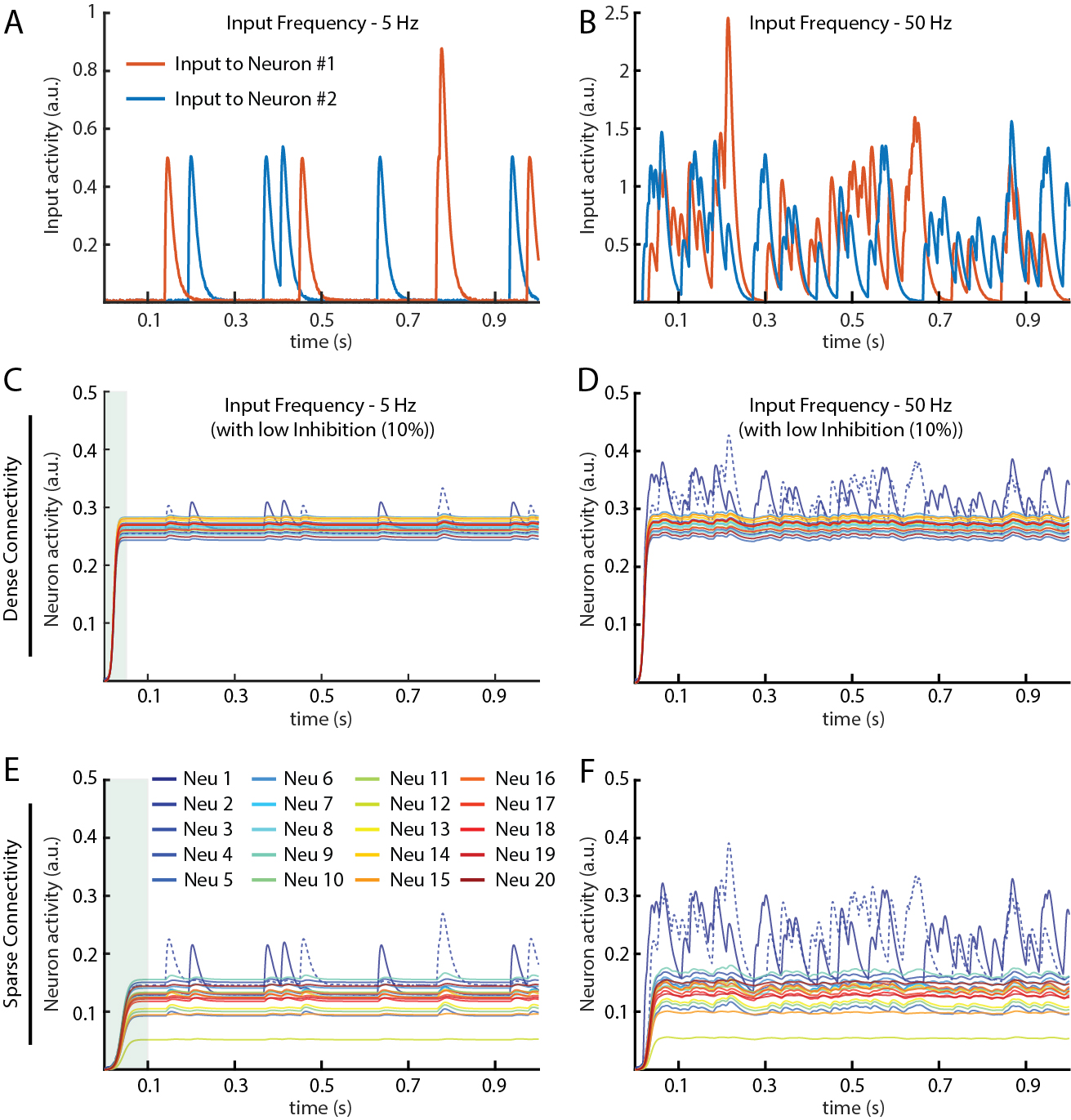}
  \caption{Neuron responses to the two given inputs for the two network configurations (dense and sparse). (\textbf{A}) The 5 Hz input, consisting of two pseudo-randomly generated input spike trains with an average frequency of 5 Hz that were convoluted using a kernel function (see Methods). The convoluted signals were fed as input to the Neurons \#1 \& \#2. (\textbf{B}) Similar to \textbf{A}, but with input spike frequencies of 50 Hz. (\textbf{C-D}) The neuron responses in the densely connected network, for the two given input frequencies. For clarity, the activity of Neuron \#2 is indicated with a dashed line. (\textbf{E-F}) The neuron responses in the sparsely connected network, for the two given inputs. In this Figure, both networks had 10\% inhibition (2 INs), an input noise of 0.1 and the initial neuron activity was 0.}
  \label{fig2}
\end{figure}

\section{Results}
\subsection{Network Activity Dynamics}
\subsubsection{Effect of LSM in bi-directional networks}

We first characterized the responses of all the neurons within the networks, with dense and sparse connectivity respectively (Figure \ref{fig1}A, B), for the two given pseudo-random inputs with average impulse frequencies of 5 Hz and 50 Hz (Input \#1 \& \#2, Figure \ref{fig2}A, B). In Figure \ref{fig2}, both the densely and the sparsely connected networks have 10\% INs (2 INs, as shown in Figure \ref{fig1}A, B) with an input noise level of 0.1 and initial neuron activity set to 0. During the initial phase of the input presentation, a gradual increase in the neuron activities was observed (shaded region, Figure \ref{fig2}A, E). Due to the recurrency within these networks, the input activity (only noise input in initial cycles (0 – 150 ms) for Input \#1) integrated over time, which caused the activity increase. Remarkably, the static leak component of the LSM normalizes the neuron activity and bring the neurons to a \emph{“steady state”\footnote{State of the neuron at which its output activity is not affected by trivial changes in the input activity.}} once a certain activity level is approached (in Figure \ref{fig2}C, for example, this occurred at a neuron activity just above 0.25). From Figure \ref{fig2} we can observe that once a steady state neuron activity is reached, only the neurons that receive direct input (Neuron \#1 \& \#2) display a sustained activity dynamics in relation to the input, whereas the other neurons displayed a substantially smaller amount of activity dynamics. Due to the density of the connections in both these network configurations, the activity dynamics of these other neurons was diminished, which was shown by the cross-correlation analysis across all the neuron responses (See Methods). The cross-correlation index (CCI), calculated across all neuron pairs for each of these configurations, was 0.98 (Figure \ref{figS1}).

\subsubsection{Effect of Inhibition}
\begin{figure}[ht] 
  \centering
  \includegraphics[width=\textwidth, center]{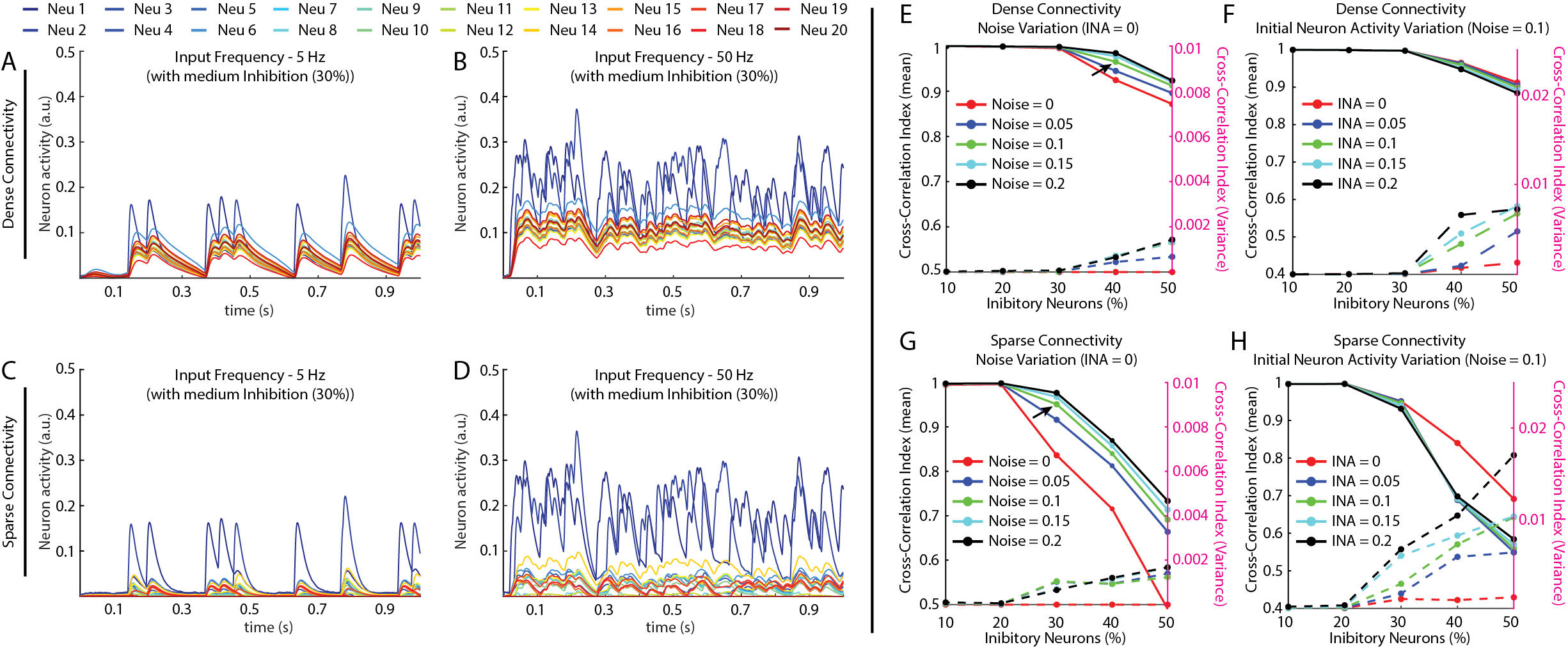}
  \caption{Neuron activity dynamics in the presence of varied inhibition and varied intrinsic properties. (\textbf{A-D}) Neuron responses for both densely and sparsely connected networks, for both given inputs (5 \& 50 Hz). These neuron responses are for network configuration with input noise = 0.1 and initial neuron activity = 0, related to the CCI index indicated by arrow in \textbf{E, G} respectively. (\textbf{E}) Cross-correlation between all neuron responses (dense connectivity, with Input \#1 (5 Hz)) for different values of input noise and inhibition. The initial neuron activity value was 0. Note: The y-axis (left) indicates the average cross-correlation index value for 25 different repetitions of each setting (see Methods), y-axis (right) indicates the variance in cross-correlation index value for 25 repetitions. This was similar for subplots \textbf{E-H}. (\textbf{F}) Cross-correlation between all neuron responses (dense connectivity) for different values of initial neuron activity and inhibition. The input noise value was 0.1. (\textbf{G}) Similar to \textbf{E}, with sparse connectivity. (\textbf{H}) Similar to \textbf{F}, with sparse connectivity.}
  \label{fig3}
\end{figure}

We next assessed the changes in network activity dynamics that could be caused by different amount of inhibition. With an increased inhibition to 30\% (i.e., the networks were configured with 6 INs), the steady state activity across all neurons, in both sparse and dense networks, was substantially lower (Figure \ref{fig3}A, C) than in the networks with 10\% inhibition (Figure \ref{fig2}). Increased inhibition would be expected to drive the neuron activity further towards zero. The LSM neuron is thresholded at zero and therefore increased inhibition would be expected to introduce non-linear dynamics into the network. This in turn would be expected to result in an increase in the activity dynamics across all the neurons of the networks, and this is also what we observed (Figure \ref{fig3}A-D). The CCI measure for each given configuration (Input noise = 0.01 \& Neuron initial activity = 0, solid green trace in Figure \ref{fig3}E, G) also indicated that the dissimilarities between the individual neuron activities increased with increased inhibition This indicated that the inhibition preserved activity dynamics in both sparsely and densely connected networks. The effect of inhibition was more pronounced in networks with sparse connectivity than with dense connectivity.

\subsubsection{Effect of Input/Neuron parameters}
We further investigated the effect of input noise level and initial neuron activity (INA) on the network activity dynamics, for different amounts of inhibition (Figure \ref{fig3}E-H). We examined the networks for five different values of input noise (0, 0.05, 0.1, 0.15 and 0.2, Noise 1-5) and INA (0, 0.05, 0.1, 0.15, 0.2, INA 1-5). For each given value of input noise and INA, we generated 25 pseudo-random repetitions (see Methods). In Figure \ref{fig3}E-H, the solid lines indicate the average CCI (y-axis on left) across these repetitions and the dotted lines indicate the variance of CCI across these repetitions (y-axis on right). \emph{Note:} The Figure \ref{fig3}E-F and Figure \ref{fig4}E-F show the results for input \#1 (5 Hz), the results for inputs \#2 (50 Hz) are presented in the supplementary materials Figure \ref{figS3}.

Figure \ref{fig3}E \& G shows that the input noise impacted the CCI for both networks. As the input noise increased, the steady state level of the neuron increased proportionally (Figure \ref{figS2}), thus affecting the dynamics in the network activity, by increasing the requirement of the amount of inhibition to create an increase in the separation in the activity between the neurons (a decrease in CCI). Conversely, for a fixed inhibition level, a greater noise caused a greater CCI value and a drop in the CCI variance, in effect a loss of separation of neuron activities. Figure \ref{fig3}F \& H illustrates the effect of INA on the activity across all the neurons in the networks. The addition of an initial neuron activity caused a change in the separation of the neuron output activity, but further changes in the INA had a negligible effect. However, increases in the INA caused a higher level of variance in CCI across a range of inhibitory amounts, indicating that the initial activity of individual neurons had a significant effect on defining the distribution of neuron output activity across the network.

\subsubsection{Effect of Conduction Delays}
\begin{figure}[ht] 
  \centering
  \includegraphics[width=\textwidth, center]{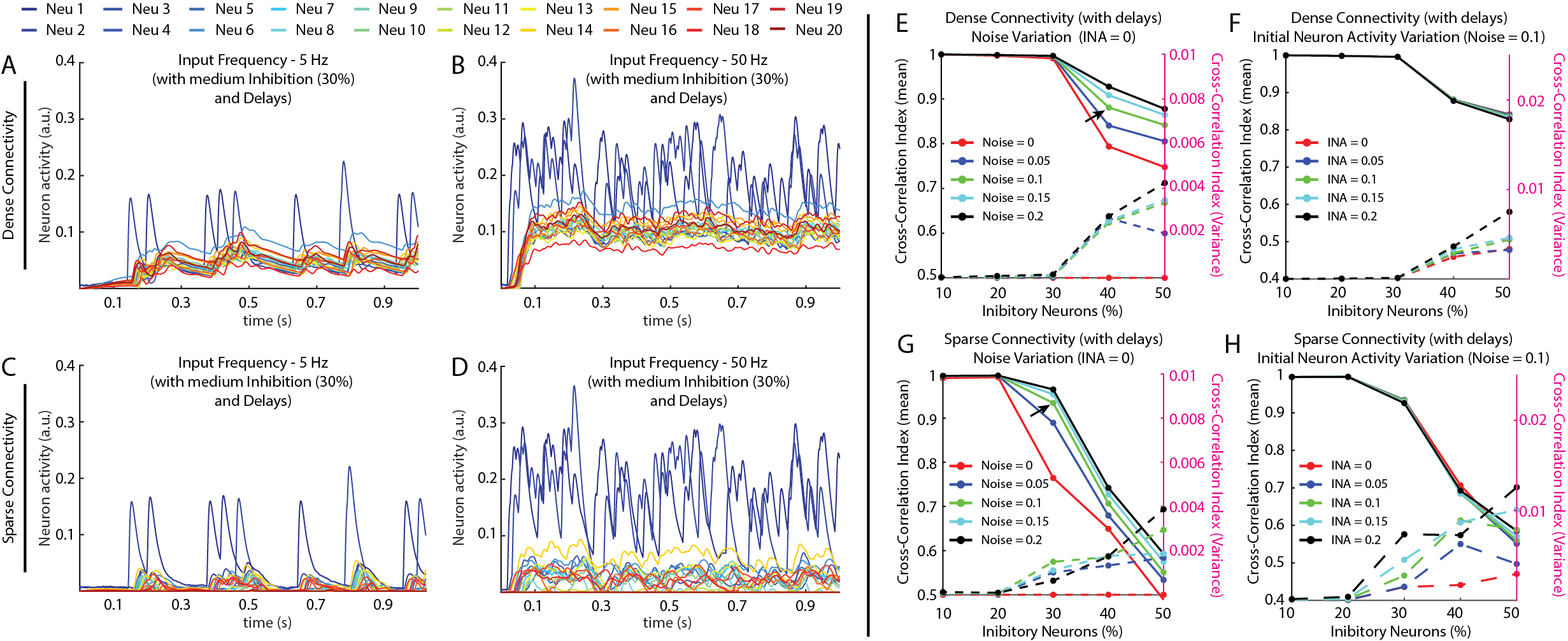}
  \caption{Neuron activity dynamics in the presence of conduction delays, for a varied inhibition and varied intrinsic properties. (\textbf{A-D}) Neuron responses for both densely and sparsely connected networks, for both given inputs (5 \& 50 Hz). These neuron responses are for network configuration with input noise = 0.1 and initial neuron activity = 0, related to the CCI index indicated by arrow in \textbf{E, G} respectively. (\textbf{E}) Cross-correlation between all neuron responses (dense connectivity, with input \#1 (5 Hz)) for different values of input noise and inhibition. The initial neuron activity value was 0. \emph{Note}: The y-axis (left) indicates the average cross-correlation index value for 25 different repetitions of each setting (see Methods), y-axis (right) indicates the variance in cross-correlation index value for 25 repetitions. This was similar for subplots \textbf{E-H}. (\textbf{F}) Cross-correlation between all neuron responses (dense connectivity) for different values of initial neuron activity and inhibition. The input noise value was 0.1.  (\textbf{G}) Similar to \textbf{E}, with sparse connectivity. (\textbf{H}) Similar to \textbf{F}, with sparse connectivity.}
  \label{fig4}
\end{figure}

We further explored if the effects of input noise, INA or inhibition could be made to alter if conduction delays between the neurons were introduced in the networks. Conduction delays seemingly induced dynamical discrepancies in activity across all the neurons in the network (Figure \ref{fig4}A-D) in contrast to the network without delays, where the activity across all the neurons converges to an attractor state based on the input signal (Figure \ref{fig3}A-D). This would be an effect of the fully connected network architecture with random weights. The effect of discrepancies across the neuron activities created by the conduction delays was expected to be more effective when the steady state activity of the neurons was near to the threshold of LSM, as this amplifies the relative differences between the neurons in the network and further results in inducing higher level of non-linearity into the network. This was reflected accordingly in the cross-correlation analysis where the CCI is lower for a given inhibition in the networks with conduction delays (Figure \ref{fig4}E-H) as compared to the networks without delays (Figure \ref{fig3}E-H). The introduction of conduction delays showed an increase in the CCI variance for densely connected networks at higher amount of inhibition (30-50\% inhibition), for different noise levels (Figure \ref{fig4}E). This variance for different noise levels indicated that the temporal effect of noise had a higher impact on the neuron output activity in the networks with conduction delays, implying a higher degree of non-linearity in the network activity dynamics compared to the networks without delays (Figure \ref{fig3}E).

\subsection{Network Dynamics}
\begin{figure}[ht] 
  \centering
  \includegraphics[width=\textwidth, center]{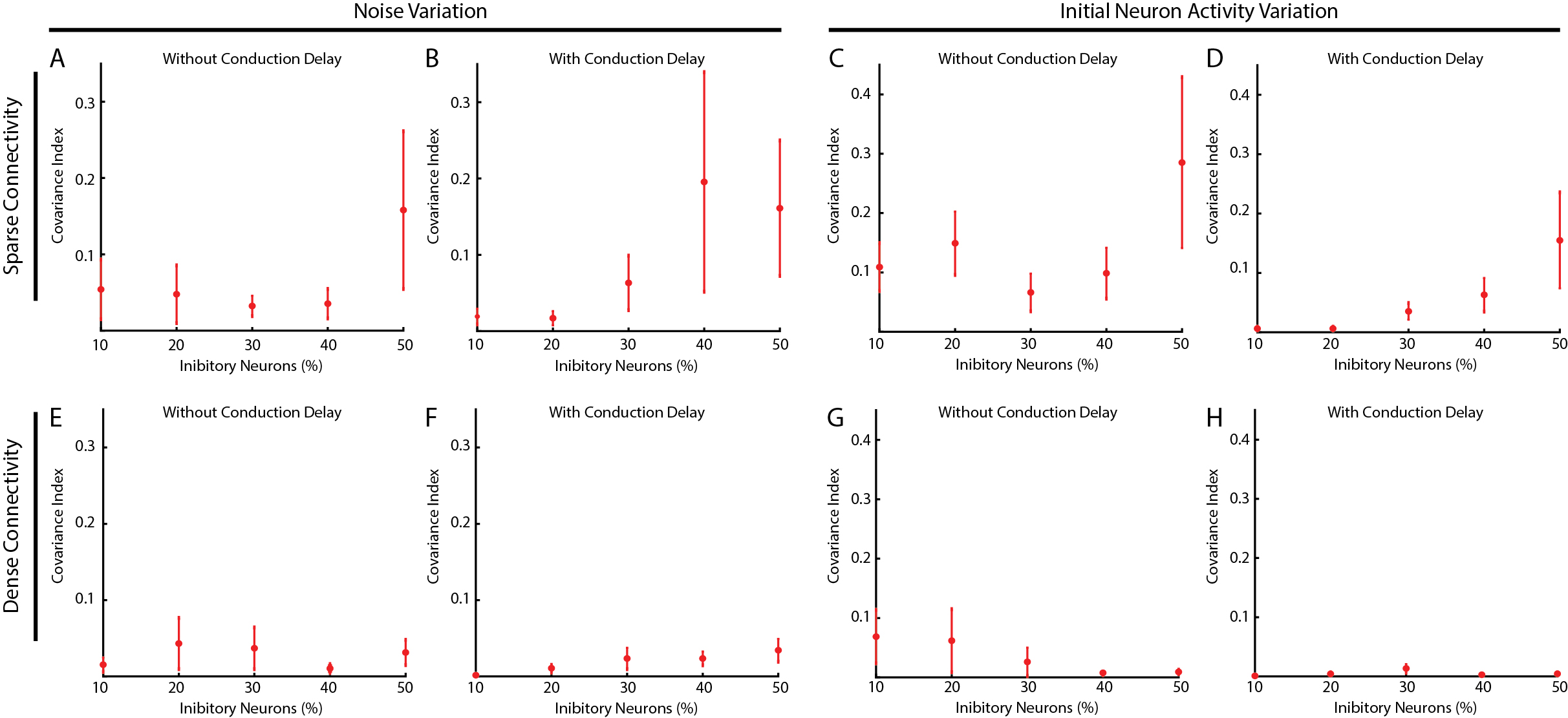}
  \caption{Analysis of network dynamics using covariance measure. The covariance index (CI) value (each data point, in each subplot) indicates the distance between covariances of neurons in a sparse/dense network, given different values of input noise/initial neuron activity.}
  \label{fig5}
\end{figure}

In the above section we explored the effect of various network configurations on the neuron activity for a given input. In this section we extend our analysis to understand if the co-activation of neurons changes for a given network configuration (input noise, INA, inhibition). We computed the covariance across all the neuron output activity and calculated the distance between covariances for different network configuration (see Methods). 
In Figure \ref{fig5}, the covariance index (CI) value (each data point, in each subplot) indicates the distance between covariances of neurons in the two network types, given different values of input noise/initial neuron activity. Figure \ref{fig5}A-D shows how the co-activation between neurons changes based on their level of input noise/INA. This variation is high in the sparsely connected networks with high amount of inhibition (30\%-50\%). In the densely connected network, the co-activation between neurons changed based on the level of input noise/initial neuron activity, but only in the networks without conduction delays (Figure \ref{fig5}E, G). Hence, in networks with conduction delays (Figure \ref{fig5}F, H) the impact of varying the configurations on causing variation in the co-activation of the neurons, was very small. 
It is clear from Figure \ref{fig5}, that sparse connectivity had a higher effect on network dynamics than dense connectivity. This also supports the argument (as mentioned in above section) that in dense networks, the steady state neuron activity is high above the threshold of LSM, preventing the neuron activity to cross the threshold, which in turn affects the network dynamics (all neurons converge to an attractor state based on the input signal).

\section{Conclusions}
\label{sec:conclusion}
The linear summation neuron model (LSM) demonstrated to be effective in keeping the total network activity at a steady state, in a fully connected recurrent neuronal network, with continuous inputs. While continually keeping the neuron activity around steady state, rather than saturating, we also showed that LSM maintains sensitivity in capturing the evolution of the input activity (Figure \ref{fig2}). This property of the LSM reflects the resting membrane potential of biological neurons \cite{rongala2021}.

The function of the thresholding function in the LSM can be related to the spiking threshold in a biological neuron. We showed that this threshold function, along with inhibition and conduction delays, to play a beneficial role for sustaining the dynamic information across all the neurons within the network. Maintaining the steady-state activity near to the threshold allowed the inhibitory input into the neuron to drive the neuron activity towards its threshold (Equation \ref{eqn2}), thereby generating non-linear dynamics in the network. Conduction delays on other hand induced discrepancies across the neurons in the network, adding an extra dimension of complexity to the non-linear dynamics in the network. Such mechanisms were demonstrated to help to avoid loss of activity dynamics in the neuron activity (and) avoiding the activity of all neurons to converge to a single attractor state (Figure \ref{fig3}, \ref{fig4}). These mechanisms would also help in learning generalized representation of the inputs, leading to a robust learning architecture.

We showed the type of network connectivity (with dense recurrency) to have significant impact on the neuron activity dynamics and network dynamics. The sparsely connected network was more sensitive to the dynamical activity changes within the input and the internal activity of the network. However, neurons in a sparsely connected network would have the disadvantage of limited integration of information across the network. The neurons in a densely connected network tend to encapsulate all the information within the network, allowing information sharing and learning generalized representations, but carrying shortcomings such as loss in information specificity. Therefore, we believe \emph{dense coding\footnote{Dense coding is when all the neurons in a network are active and their combined activity is used to encode each context.}}  in a sparsely connected network will be potentially an advantageous tradeoff between information specificity and generalization of information representation \cite{spanne2015}. 

Considering a dense coding scenario (where the activity across all neurons was used to learn a given context) the internal state of the neurons would play a crucial role. The combined internal state of neurons in a network will define the network state, which in turn become the reference by which the input would be learnt. In this work, with the covariance analysis (Figure \ref{fig5}), we showed that the covariance between neurons would vary depending on the specific input/neuron parameters and the network connectivity. Therefore, we argue that state of the neuron/network is dynamically changing and it will become a defining factor for how information is represented within the network.

\begin{ack}
This work was supported by the EU Grant FET 829186 ph-coding (Predictive Haptic COding Devices In Next Generation interfaces), the Swedish Research Council (Project Grant No. K2014-63X-14780-12-3).
\end{ack}

\medskip

{
\small


}

\newpage
\appendix
\renewcommand\thefigure{\thesection.\arabic{figure}}    
\section{Appendix}
\setcounter{figure}{0}

\begin{figure}[ht] 
  \centering
  \includegraphics[width=\textwidth, center]{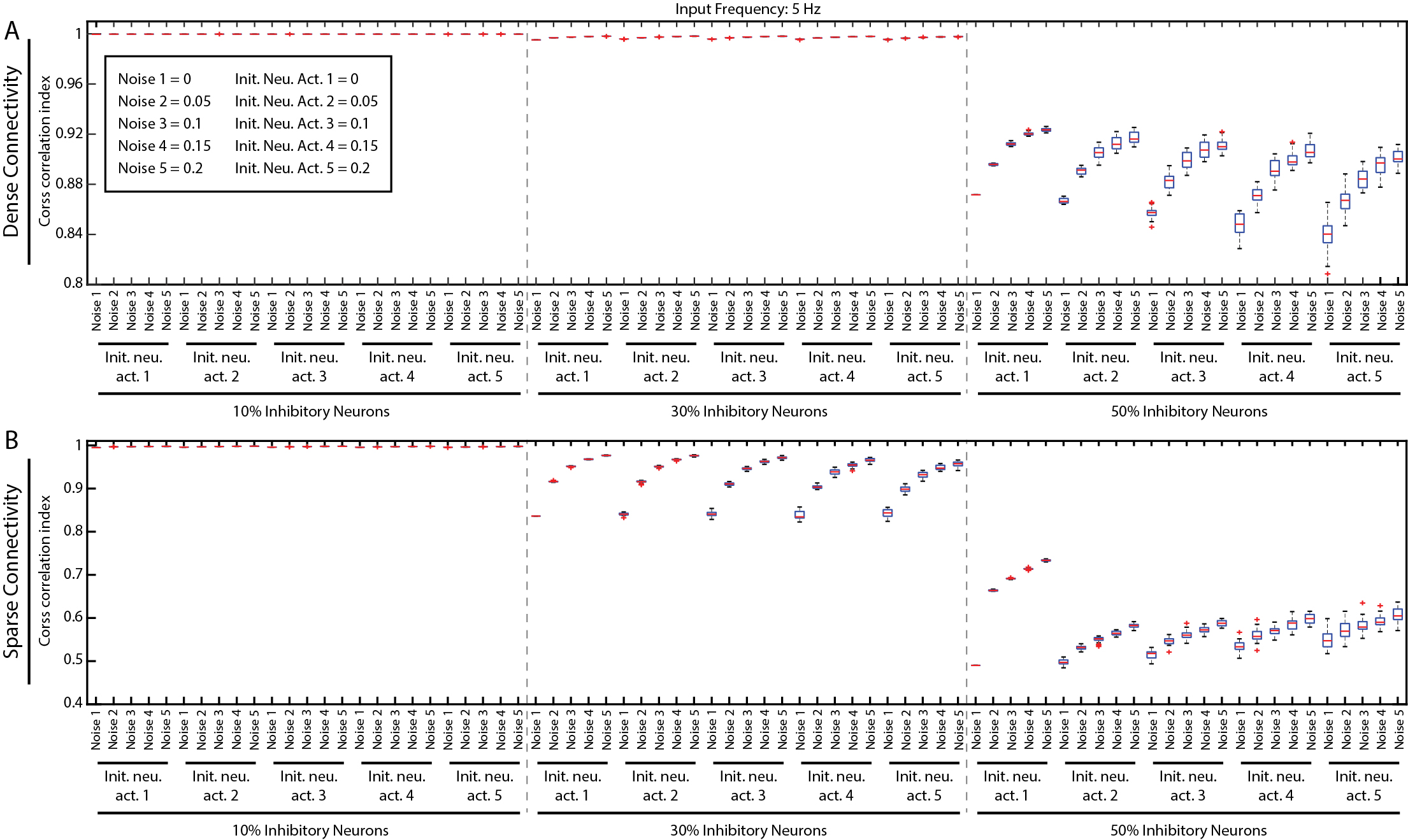}
  \caption{(\textbf{A}) Cross-correlation between all neuron responses (dense connectivity) for different values of input noise, initial neuron activity and inhibition. The boxplot was made across the correlation measure for 25 different repetitions of the same network setting (input noise, initial neuron activity, inhibition). \textbf{B} Similar to \textbf{A}, with sparse connectivity.}
  \label{figS1}
\end{figure}

\newpage
\begin{figure}[ht] 
  \centering
  \includegraphics[width=\textwidth, center]{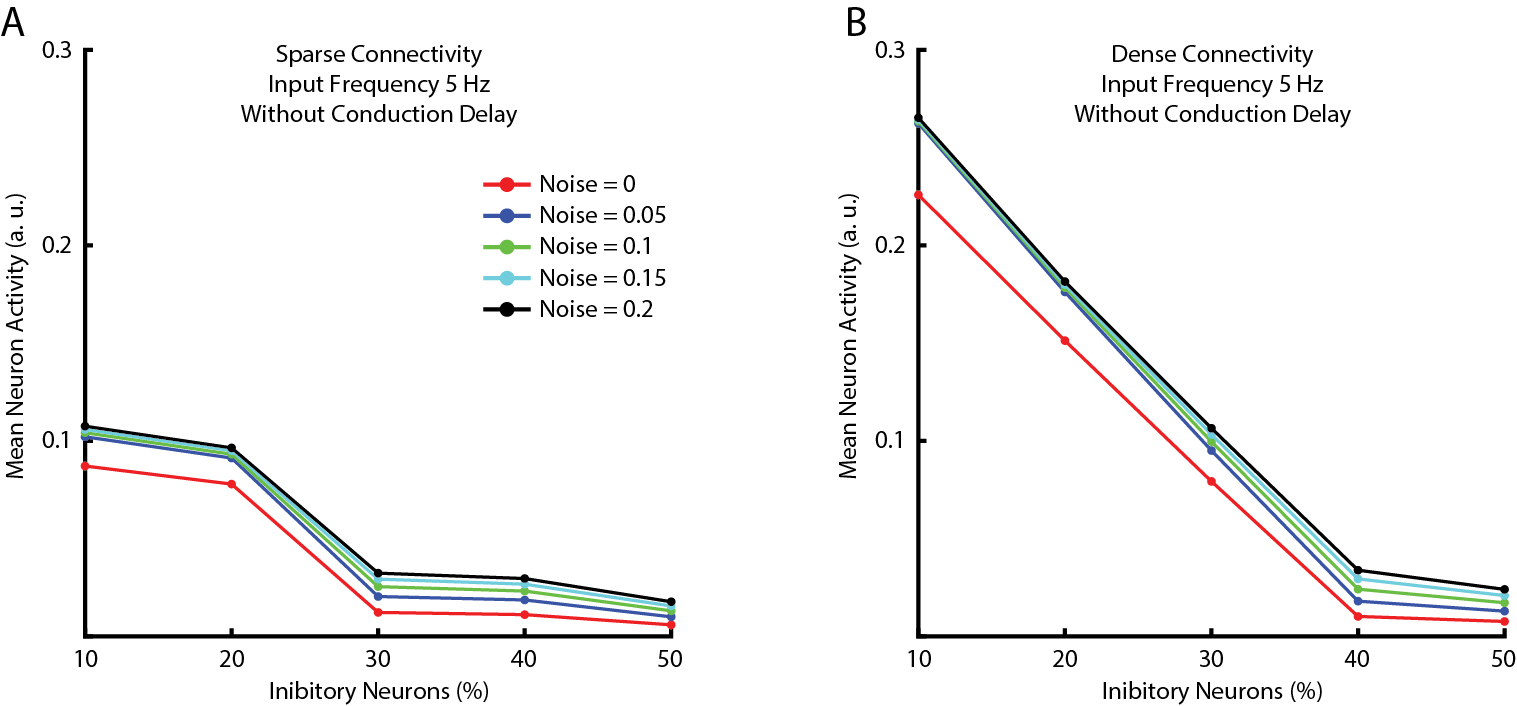}
  \caption{Average neuron activity. (\textbf{A}) Average neuron activity in a network with sparse connectivity, for different input noise and a given input of 5 Hz (input \#1). (\textbf{B}) Similar to \textbf{A}, with densely connected network.}
  \label{figS2}
\end{figure}

\newpage
\begin{figure}[ht] 
  \centering
  \includegraphics[width=\textwidth, center]{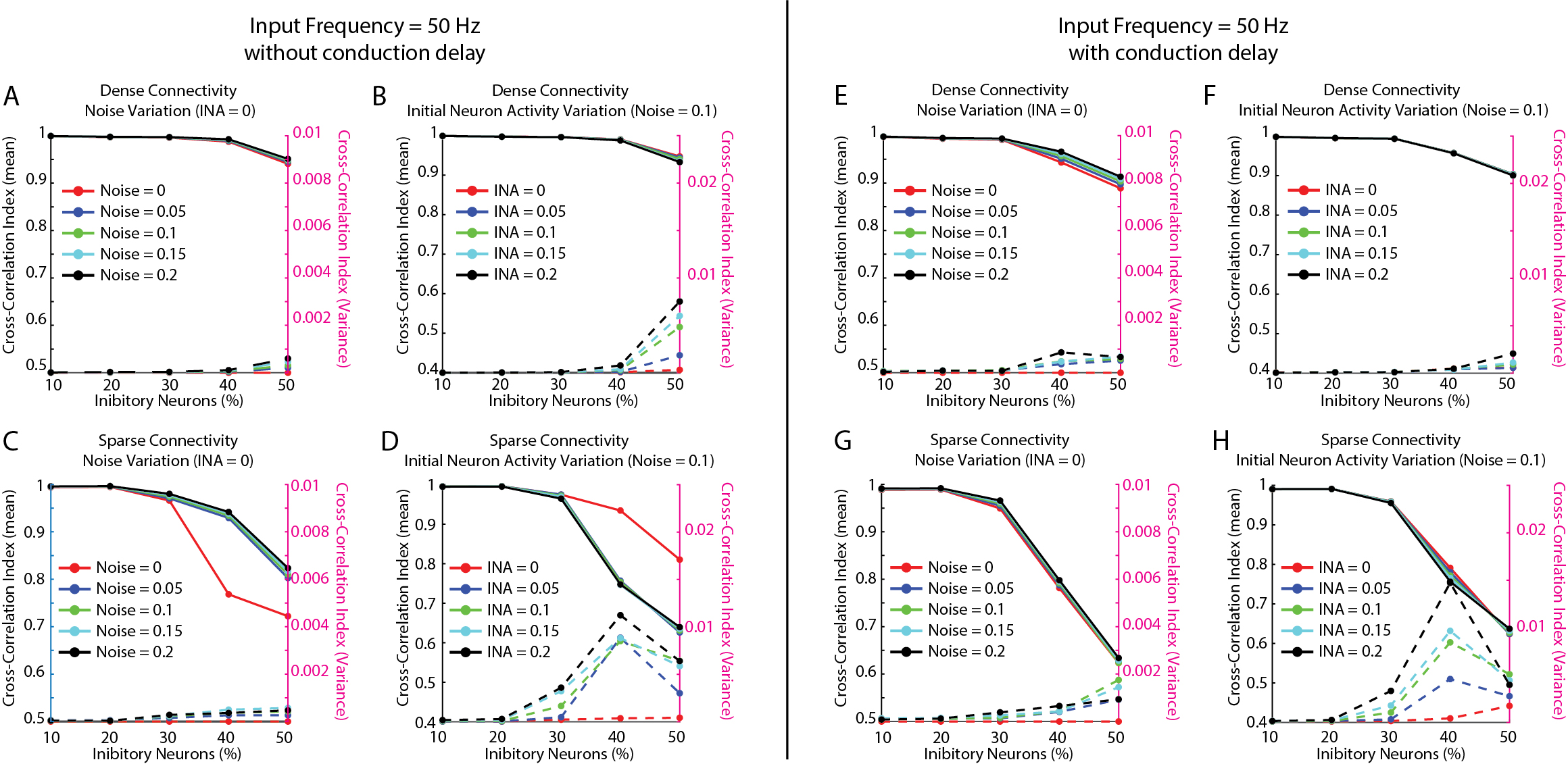}
  \caption{(\textbf{A}) Cross-correlation between all neuron responses (dense connectivity, with input \#2 (50 Hz)) for different values of input noise and inhibition. The initial neuron activity value was 0. Note: The y-axis (left) indicates the average cross-correlation index value for 25 different repetitions of each setting (see Methods), y-axis (right) indicates the variance in cross-correlation index value for 25 repetitions. This was similar for subplots \textbf{A-H}. (\textbf{B}) Cross-correlation between all neuron responses (dense connectivity) for different values of initial neuron activity and inhibition. The input noise value was 0.1.  (\textbf{C}) Similar to \textbf{A}, with sparse connectivity. (\textbf{D}) Similar to \textbf{B}, with sparse connectivity. \textbf{E-H} Similar to \textbf{A-D}, with the networks in the presence of conduction delays.}
  \label{figS3}
\end{figure}

\end{document}